\documentclass[10pt,twocolumn,aps,manuscript,a4paper]{revtex4}
\usepackage[T1]{fontenc}
\usepackage[latin1]{inputenc}
\usepackage{graphics}

\begin{document}

\title{A dynamical approach of the microcanonical ensemble}

\author{Xavier Leoncini}
   \email{leoncini@cims.nyu.edu}
   \affiliation{Courant
       Institute of Mathematical Sciences,
       New York University,
       251 Mercer St.,
       New York NY 10012 USA}

\author{Alberto Verga}
   \email{Alberto.Verga@irphe.univ-mrs.fr}
   \affiliation{
      Institut de Recherche sur les Ph\'enom\`enes Hors \'Equilibre,
      49, rue F. Joliot-Curie, BP 146,
      13384 Marseille, France}

\begin{abstract}
An analytical method to compute thermodynamic properties of a given
Hamiltonian system is proposed. This method combines ideas of both
dynamical systems and ensemble approaches to thermodynamics, providing
de facto a possible alternative to traditional Ensemble methods.
Thermodynamic properties are extracted from effective motion
equations. These equations are obtained by introducing a general
variational principle applied to an action averaged over a statistical
ensemble of paths defined on the constant energy surface. The method
is applied first to the one dimensional \( \beta  \)-FPU chain and to
the two dimensional lattice \( \varphi ^{4} \) model. In both cases
the method gives a good insight of some of their statistical and
dynamical properties.
\end{abstract}
\maketitle

The problem raised by Clausius and the second principle found its
answer with Boltzmann and the rise of equilibrium statistical
physics\cite{Gal94,Max90}. An essential point in the theory is related
to the law of large numbers, which ensures that fluctuations around
mean values of the thermodynamic quantities are negligible \cite{LL}.
The concept of \emph{ensembles} is introduced, as for instance the
microcanonical ensemble for isolated systems, and their associated
measures are used to average. Developments within the
Ensemble framework have generalized the use of various techniques such
as perturbation expansions, mean field approximation, or
renormalization group \cite{Lebellac} and greatly improved our
understanding of phase transitions phenomena (see for instance the
review \cite{Pelissetto00} and references therein). However, the
computation of thermodynamic properties for a given Hamiltonian system
remains in general inextricable.

The purpose of this Letter is to introduce an analytical approach of
the thermodynamic limit and provide an alternative to classical
techniques. This method relies on the large size limit and the
universality of trajectories (good ergodic and mixing properties are
assumed). We define an ensemble of paths drawn on the energy surface
and compute thermodynamic variables through averaged equations of
motion. This approach applies to systems at equilibrium, and proves to
be very successful in the chosen examples. Note that, the ensemble
averaging implies a large time limit before the large system limit,
but we invert the order of these two limits.

\begin{figure}
{\par\centering \resizebox*{7cm}{!}{\includegraphics{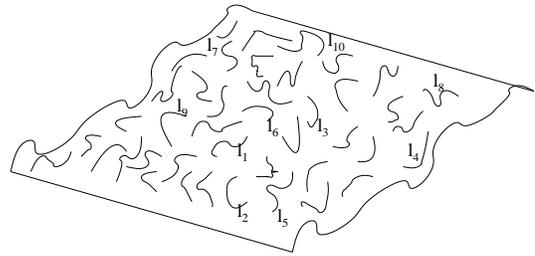}}
\par}
  \caption{Representation of the constant energy surface. Different
  labeled paths are drawn on it.
  \label{Fig1}}
\end{figure}

Let us identify a set of trajectories on the hypersurface defined by
the microcanonical measure in the phase-space  by a set of labels \(
{\ell } \), which may be initial conditions for instance. The
thermodynamic state does then not depend on these labels (this
property permits the introduction of the ensemble averaging). In the
same spirit, we consider a family of paths \( q_{\ell }(t) \) (we
noted explicitly the time \( t \) and label \( {\ell } \) dependences)
drawn on the constant energy surface (see Fig. \ref{Fig1}). To each
path we associate a Lagrangian \( L(q_{\ell },\dot{q}_{\ell }) \)
where the dot denotes time derivative and the corresponding action \(
A=\int dt\, L \). The basis of the proposed method relies on the
following claim: since the thermodynamic state is label independent,
we can average the Lagrangian over the labels, and apply the
variational principle on the \emph{mean dynamical system}:
\begin{equation}
  \label{A}
  \langle \delta A\rangle =\delta \langle A\rangle =\delta \int dt\,
  \langle L\rangle =0\: ,
\end{equation}
(where \( \langle \cdots \rangle  \) denotes averaging over the
labels). The second equality in (\ref{A}) is imposed as a
compatibility condition at equilibrium and defines a smooth path as
the average of a flow of paths of the original system. We note that
after the average is performed, trajectories and points related to the
\emph{mean dynamical system} must already comprise some information on
the thermodynamic state, hence we shall refer to the resulting motion
equations as \emph{thermodynamic motion equations}.
Let us now consider Hamiltonian systems of the type \(
H=\mathbf{p}^{2}/2+V(\mathbf{q}) \), namely quadratic in momentum and
with separated conjugated variables. Microcanonical statistics leads
to a linear relation between the mean kinetic energy \( \mbox
{MC}\mathbf{p}^{2}/2 \) and the temperature \( T \) (\( \mbox
{MC}\cdots  \) stands for microcanonical averaging) \cite{Pea85} and
predicts that the momentum is Gaussian with each component \( p_{i} \)
independent and a variance proportional to the temperature \( \mbox
{MC}p_{i}^{2}\sim T \). In the canonical ensemble this results in a
trivial factorization of the partition function, all the complexity
being included in the potential \( V \). The present approach uses
this Gaussian property and reverse the usual argument to pass from
time averaging to ensemble averaging: at thermal equilibrium, we
interpret \( p \) as being a Gaussian stochastic process on the labels
and get thermodynamic quantities from the mean dynamical system. We
now propose a possible implementation of these ideas.

We consider a lattice (in dimension \( D=1,2 \)) of \( N \) sites with
coordinates \( x_{i},i=1,\cdots ,N \). At each site \( i \) is placed
a particle, in general coupled to its neighbors, having momentum \(
p_{i} \) and conjugate coordinate \( q_{i} \). We take units such that
the lattice spacing, the Boltzmann constant, and the mass are equal to
one. Since \( p_{i} \) is Gaussian, we choose to represent it as a
superposition of random phased waves:
\begin{equation}
  \label{p}
  p_{i}=\sum _{k=0}^{Nk_{0}}\dot{\alpha }_{k}\cos (kx_{i}+\phi _{k})\: ,
\end{equation}
where the wavenumber \( k \) is in the reciprocal lattice (an integer
multiple of \( k_{0}=2\pi /N^{(1/D)} \)), the wave amplitude is \(
\dot{\alpha }_{k} \), and its phase \( \phi _{k} \) is uniformly
distributed on the circle. The momentum set is labeled, using
(\ref{p}), with the set of phases \( \ell \equiv \{\phi _{k}\} \).
This equation can also be interpreted as a change of variables, from
\( p \) to \( \alpha  \), with constant Jacobian (the change is linear
and we chose an equal number of modes and particles). Besides, if the
total momentum is conserved, we choose to take \( \dot{\alpha }_{0}=0
\). As the variance of \( p_{i} \) is fixed, we shall assume that the
\( \dot{\alpha }_{k} \) are all of the same order (we need a large
number of relevant modes for the center-limit theorem to apply). Using
the relation \( \langle p_{i}^{2}\rangle =\sum \dot{\alpha }_{k}^{2}/2
\) (we average over the random phases) and imposing that at
equilibrium the fluctuations are small, we write \( \langle
p_{i}^{2}\rangle \approx T \) and obtain \( \dot{\alpha
}^{2}_{k}\approx O[(T/N^{})] \) (we call this relation the Jeans
condition \cite{Jea24}). We shall see in the examples that for this
scaling in \( N \) for \( \dot{\alpha } \) and the short range
interaction, the mean dynamical system becomes a set of oscillators
with mean-field type interactions and a kind of Jeans spectrum. The
coordinate variables associated with the representation of momenta
(\ref{p}) are
\begin{equation}
  \label{q}
  q_{i}=\alpha _{0}+\sum _{k=k_{0}}^{Nk_{0}}\alpha _{k}
    \cos (kx_{i}+\phi _{k})\: .
\end{equation}
Note that this equation supposes true the relation \(
p_{i}=\dot{q}_{i} \). The equilibrium state is constructed from the
averaged Lagrangian \( {\mathcal{L}}=\langle L\rangle /N \), the
condition that the paths belong to the energy surface \(
e(T)=E/N=\langle H\rangle /N \), and the Jeans condition which fixes
the temperature from the averaged kinetic energy. We in fact applied a
version of this method to the Kosterlitz-Thouless phase transition in
the \( XY \) model \cite{Leo98}.

We shall start to test this approach with the generic case of a chain
of coupled harmonic oscillators. The Hamiltonian writes \( H=(1/2)\sum
_{i}[p_{i}^{2}+(q_{i+1}-q_{i})^{2}] \). Using the expressions
(\ref{p}) and (\ref{q}) we compute the averaged Lagrangian \(
{\mathcal{L}}=(1/4)\sum _{k}[\dot{\alpha }_{k}^{2}-\omega
_{0k}^{2}\alpha _{k}^{2}] \) where \( \omega _{0k}^{2}=4\sin ^{2}k/2
\) and extremize the action to obtain the thermodynamic motion
equations \( \ddot{\alpha }_{k}=-\omega _{0k}^{2}\alpha _{k} \).
Equilibrium is imposed by the Jeans condition which gives \( \alpha
_{k}^{2}=2T/N\omega _{0k}^{2} \) and leads to the thermodynamic
function \( e(T)=T \).

\begin{figure}
{\centering \resizebox*{7cm}{!}{\includegraphics{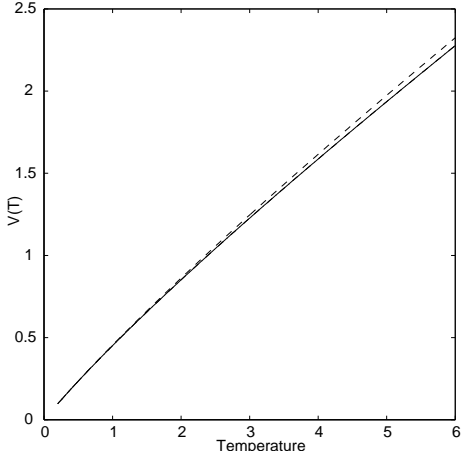}}}
\caption{Potential energy versus temperature for \(\beta=0.1\), the
dashed line corresponds to the result obtained   by
\protect\cite{Liv87}, while the solid line corresponds to equation
\ref{V_{F}PU_{f}inal}. \label{Fig2}}
\end{figure}

We consider now the Fermi-Pasta-Ulam problem of a one-dimensional \(
\beta  \)-FPU chain of oscillators. The thermodynamics of this model
has been exactly computed within the canonical ensemble \cite{Liv87}.
The Hamiltonian reads \( H=(1/2)\sum
_{i}[p_{i}^{2}+V(q_{i+1}-q_{i})+V(q_{i}-q_{i-1})] \), where \(
V(x)=x^{2}/2+\beta x^{4}/4 \). The averaged potential energy density
\( v=\langle V\rangle /N \) is then
\begin{equation}
  \label{v}
  v=\sum _{k}\left( s_{k}^{2}\alpha _{k}^{2}+
    {\frac{3}{2}}\beta s_{k}^{4}\alpha _{k}^{4}\right) +
    3\beta \sum _{k\ne k'}s_{k}^{2}s_{k'}^{2}
    \alpha _{k}^{2}\alpha _{k'}^{2}\: ,
\end{equation}
where we used the notation \( s_{k}=\sin (k/2) \). It is worth
noticing that by considering \( \alpha  \) as the dynamical variables
of a new system, the interaction \( v \) is of the mean-field type as
the second term in (\ref{v}) involves interaction between all \(
\alpha _{k} \) oscillators. The thermodynamic motion equations are
obtained as before from the Lagrangian density \(
{\mathcal{L}}(\dot{\alpha },\alpha )=\sum _{k}[\dot{\alpha
}_{k}^{2}/2-v(\alpha )] \):
\begin{equation}
  \label{Qfpu}
  \ddot{\alpha }_{k}=-\omega _{0k}^{2}\left[
  1+Q-3\beta s_{k}^{2}\alpha _{k}^{2}\right] \alpha _{k},\:
  Q=6\beta \sum _{k}s_{k}^{2}\alpha _{k}^{2}\: ,
\end{equation}
where \( Q \) is a mean-field (intensive) variable. Its fluctuations
at equilibrium are of order \( {\mathcal{O}}(1/\sqrt{N}) \). Hence we
consider it in (\ref{Qfpu}) as a constant. This is the large system
limit \( N\rightarrow \infty  \) taken before the \( t\rightarrow
\infty  \) limit. In this approximation (\ref{Qfpu}) describes a set
of uncoupled oscillators. Moreover, as we verify \emph{a posteriori},
to the same order of approximation we neglect the third term in the
brackets (it is smaller than the \( Q \) term, by a factor \(
{\mathcal{O}}(1/N) \)). We then obtain a simple linear wave equation
with a dispersion relation \( \omega _{k}^{2}=\omega _{0k}^{2}(1+Q) \)
\cite{Ala95}. The Jeans condition gives \( \alpha _{k}^{2}\omega
_{k}^{2}\approx 2T/N \), and allows an estimation of the neglected
term \( s_{k}^{2}\alpha _{k}^{2}\approx T/[2N(1+Q)] \). Using now the
dispersion relation, the Jeans condition, and the definition of \( Q
\), we obtain \( Q(1+Q)/3\beta =T \) and the function \( Q=Q(T) \).
Since \( v=(2Q+Q^{2})/12\beta  \) we finally get the thermodynamic
relation
\begin{equation}
  \label{V_{F}PU_{f}inal}
  v=\frac{T}{4}+\frac{\sqrt{1+12\beta T}-1}{24\beta }\: .
\end{equation}
This result is compared with the canonical one (\cite{Liv87}) in
Fig.~\ref{Fig2}. The two results are in very good agreement and we
speculate it is exact.

For the last example, we consider a two dimensional system (\( D=2 \))
exhibiting a second order phase transition, the so called dynamical
lattice \( \varphi ^{4} \) model studied in \cite{Cai98}. This model
is defined by the Hamiltonian
\begin{equation}
  \label{H4}
  H=\sum _{i=1}^{N}\left( \frac{p_{i}^{2}}{2}-
  \frac{m^{2}}{2}q_{i}^{2}+\frac{\lambda }{4!}q_{i}^{4}\right) +
  \frac{J}{2}\sum _{\langle i,j\rangle }(q_{i}-q_{j})^{2}\: ,
\end{equation}
where \( m \) and \( \lambda  \) are real parameters, \( J=1 \) is the
coupling constant, and \( \langle i,j\rangle  \) denotes the summation
over the close neighbors on a square lattice. In contrast with the
other cases, where the \( k=0 \) mode \( \alpha _{0} \) was a free
parameter, in this example it is relevant and corresponds to the
average of \( q \) which is proportional to the magnetization of the
system and is independent of time at equilibrium. The computation of
the averaged potential energy is very similar to the \( \beta  \)-FPU
case:
\begin{eqnarray}
  v & = & \frac{1}{4}\sum _{k}\left(
    \omega _{0k}^{2}-m^{2}+\frac{\lambda }{2}\alpha _{0}^{2}\right)
    \alpha _{k}^{2}+\frac{\lambda }{32}\sum _{k\ne k'}
    \alpha _{k}^{2}\alpha _{k'}^{2}\: \nonumber \\
    &  & +\frac{\lambda }{64}\sum _{k}\alpha _{k}^{4}-
    \frac{m^{2}}{2}\alpha _{0}^{2}+\frac{\lambda }{4!}
    \alpha _{0}^{4}\: .
    \label{v_{p}hi4}
\end{eqnarray}
The thermodynamic motion equations are
\begin{eqnarray}
  0 & = & \alpha _{0}\left( \frac{\lambda }{6}\alpha ^{2}_{0}+
  \frac{\lambda }{4}Q-m^{2}\right) \label{phi0eq} \\
  \ddot{\alpha }_{k} & = & -\left( \omega ^{2}_{0k}-m^{2}+
  \frac{\lambda }{2}\alpha ^{2}_{0}+\frac{\lambda }{4}Q\right)
  \alpha _{k}+\frac{\lambda }{8}\alpha ^{3}_{k}\: ,
  \label{alpha_{p}hieq}
\end{eqnarray}
where \( \omega _{0k}^{2}=4(\sin ^{2}k_{x}/2+\sin ^{2}k_{y}/2) \) is
the free harmonic frequency (\( k=(k_{x},k_{y}) \) is now a vector in
the plane), and \( Q=\sum \alpha _{k}^{2} \) is an intensive variable.

Equation (\ref{phi0eq}) has multiple solutions in \( \alpha _{0} \)
depending on the temperature through \( Q=Q(T) \). Since \( \alpha
_{0}=\langle q_{i}\rangle  \) is the order parameter, we anticipate
the existence of a phase transition in the thermodynamic state.
Indeed, the only solution is \( \alpha _{0}=0 \) for \(
Q>Q_{*}=4m^{2}/\lambda  \) but for \( Q<Q_{*} \) other solutions with
finite values of the order parameter exist:
\begin{equation}
  \label{LowT_{m}ag}
  \alpha _{0}^{2}=\frac{6m^{2}}{\lambda }-\frac{3}{2}Q=\frac{3}{2}(Q_{*}-Q)\: .
\end{equation}
To solve (\ref{alpha_{p}hieq}), we neglect the \( \alpha _{k}^{3} \)
term, as we did for the \( \beta  \)-FPU case (large system limit) and
obtain a wave equation with the dispersion relation: \( \omega
_{k}^{2}=\omega _{0k}^{2}+\Omega ^{2}(T) \). \( \Omega (T) \) is given
by,
\begin{eqnarray}
  \Omega ^{2} & = & \frac{\lambda }{2}(Q_{*}-Q),\quad Q<Q_{*}
  \label{Omega_{l}ow} \\
  \Omega ^{2} & = & \frac{\lambda }{4}(Q-Q_{*}),\quad Q>Q_{*}\: .
  \label{Omega_{h}igh}
\end{eqnarray}
where we used the definition of \( Q_{*} \) and
Eq.~(\ref{LowT_{m}ag}). We notice that \( \Omega  \) is the frequency
corresponding to the absent \( k=0 \) mode.

Given the Jeans condition: \( \alpha _{k}^{2}=2T/(N\omega _{k}^{2}) \)
we notice that as long as \( \Omega \ne0  \) (\( Q\ne Q_{*} \)) the \(
\alpha _{k}^{3} \) term is \emph{a posteriori} negligible, we can
therefore expect results obtained with this approximation to be
accurate everywhere but at (near) the transition. Another consequence
of this condition is that most of the modes have comparable amplitudes
for \( \Omega \ne0  \), which then implies a Gaussian-like distribution
for \( q \). At the same order of approximation, in the thermodynamic
limit, we can compute \( Q(T) \), identifying it to a Riemann
integral. Using the \( \alpha _{k} \) given by the Jeans spectrum we
obtain the following implicit equation for \( Q(T) \),
\begin{equation}
  \label{Implicit_{p}hi4}
  Q(T)=\frac{T}{\pi }aK(a),\quad a=\frac{4}{4+\Omega ^{2}(T)}\: ,
\end{equation}
where \( K(a) \) is the complete elliptic integral of the first kind
\( \int ^{\pi /2}_{0}d\theta /\sqrt{1-a^{2}\sin ^{2}\theta } \).

\begin{figure}
{\par\centering \resizebox*{7cm}{!}{\includegraphics{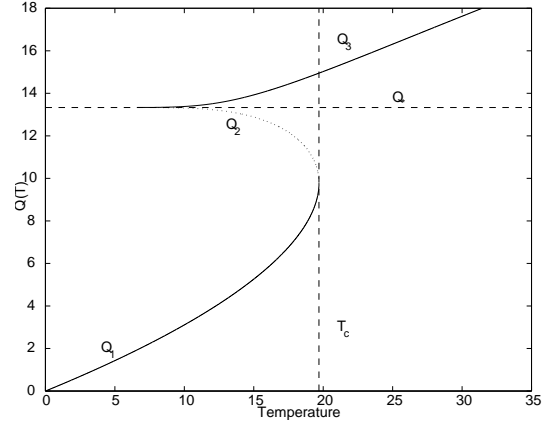}}
\par}

\caption{Solutions of the implicit equation (\ref{Implicit_{p}hi4}),
for \(m^{2}=2\) and \(\lambda=0.6\). We notice that depending on the value of \(T\) up to three different solutions for \(Q\) exists giving rise to three different branches respectively labelled \(Q_1 \), \(Q_2 \), \(Q_3 \). The critical temperature \(T_c\) is identified.  \label{Fig3}}
\end{figure}

The solutions of (\ref{Implicit_{p}hi4}) has three branches, plotted
on Fig.~\ref{Fig3}. We choose the same values \( m^{2}=2 \) and \(
\lambda =0.6 \) as the ones used in \cite{Cai98}. We first notice the
existence of a special temperature close to \( 19.69 \) localizes the
phase transition temperature. Two branches are below \( Q_{*}=40/3 \),
and result from the expression (\ref{Omega_{l}ow}) used in the
implicit equation, the third one (on the top of the figure) corresponds
to the expression (\ref{Omega_{h}igh}). The fact that the wave form
solutions are not valid for \( Q=Q_{*} \) translate in the divergence
of \( K(a) \), since as \( Q\rightarrow Q_{*} \), \( a\rightarrow 1
\). The divergence is logarithmic, and then for sufficiently small \(
T \), a solution around \( Q_{*} \) of (\ref{Implicit_{p}hi4}) always
exists; resulting in the two upper branches being asymptotic to the \(
Q(T)=Q_{*} \) curve as \( T \) goes to \( 0 \).

\begin{figure}
{\par\centering \resizebox*{7cm}{!}{\includegraphics{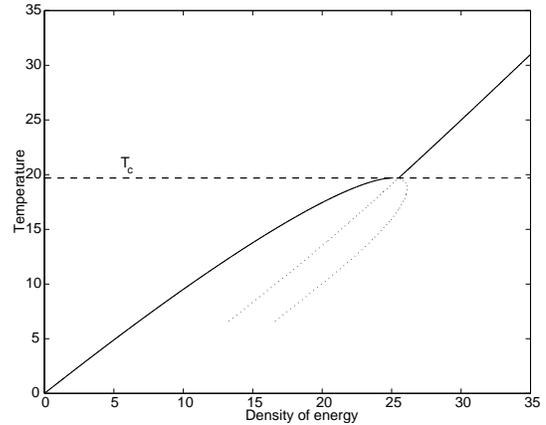}}
\par}

\caption{Temperature versus density of energy for \(m^{2}=2\)   and
\(\lambda=0.6\). The three different branches resulting from the solutions of \( Q \) presented in Fig.~\ref{Fig3} are represented. The
physical line is the upper one. The transition is clearly identified
with \(T_{c} \approx 19.64\) and \(\epsilon_{c} \approx 25.07\).
\label{Fig4}}
\end{figure}

In order to select one branch from another we compute their respective
density of energy. According to equations (\ref{v_{p}hi4}) and
(\ref{phi0eq}), we have two different expressions for the density of
energy,
\begin{equation}
  \label{en_{h}igh}
  e(T)=\left\{ \begin{array}{ll}
  T-\frac{3m^{4}}{2\lambda }+\frac{3\lambda }{32}Q^{2} & ,
  \quad Q<Q_{c}\\
  T-\frac{\lambda }{32}Q^{2} & ,
  \quad Q>Q_{c}\:
  \end{array}\right. .
\end{equation}
The results are plotted as the temperature versus the density of
energy in Fig. \ref{Fig4} in analogy to the results presented in
\cite{Cai98}. The physical relevant solution is the one whose energy
is the smaller for a given temperature, which translate in the upper
line in the figure. The transition is then identified at a density of
energy \( e_{c}=25.07 \), whose corresponding temperature is \(
T_{c}=19.69 \). These results are in good agreement with the one
predicted by numerical simulations \cite{Cai98}: respectively \(
T_{c}=17.65 \) and \( e_{c}=21.1 \). Using equation
(\ref{LowT_{m}ag}), we have also access to the square of the
magnetization and there is also a good quantitative agreement with the
numerical results. The discontinuous behavior of the magnetization at
the transition is although surprising. This behavior was also observed
numerically in \cite{Cai98}, and is explained by noticing that the
true order parameter is \( \langle |q|\rangle  \) and not \( |\langle
q\rangle | \). In the present case, we can also wonder whether this
behavior is due to the \( N\rightarrow \infty  \) limit taken before
the \( t\rightarrow \infty  \). Indeed the neglected terms are
relevant at the transition, and only become negligible around the
transition after the \( N\rightarrow \infty  \) limit, which may
affect the nature of the observed transition. However this behavior
may also find its origin in the choice of writing the momentum as a
superposition of \( N' \) random phased waves (motivated by the
solutions of the linearized equations of motion) equal to the number
of degrees of freedom \( N \). Writing the momentum with the number of
modes \( N' \) being a growing unbounded function of \( N \) is
sufficient to obtain a Gaussian process. Another representation may
then be appropriate to tackle the transition region and for instance in
\cite{Leo98} a high temperature approach was used to compute the
critical temperature.

To conclude we point out that the thermodynamic motion equations method
allowed us to compute the macroscopic properties of coupled nonlinear
oscillator systems in one and two dimensions. Quantitative agreement
with exact or numerical results of these quantities is obtained.
Moreover the phase transition for the \( \varphi ^{4} \) model is
detected and a good estimate of the critical energy and temperature
are given even though we approximatively solved the thermodynamic
motion equations. We expect that this method will be successful for
other systems and speculate that the actual solving of the exact
thermodynamics motion equation should lead to an exact thermodynamic
limit. We believe it may also be possible to extend the scope of the
method to systems out of equilibrium and describe their macroscopic
evolution with the thermodynamic motion equations.

\acknowledgments{We thanks fruitful discussions with S. Ruffo.}

\end{document}